\documentclass[11pt]{article}
\usepackage{fullpage,epsf,amssymb,amsthm,amsfonts,amsmath,latexsym, graphicx,array,extarrows,mathtools,mathstyle,mathrsfs,slashed}
\usepackage[normalem]{ulem}

\usepackage[margin=1in]{geometry}

\newcommand{\overbar}[1]{\mkern 1.5mu\overline{\mkern-1.5mu#1\mkern-1.5mu}\mkern 1.5mu}

\let\oldsqrt\sqrt
\def\sqrt{\mathpalette\DHLhksqrt}
\def\DHLhksqrt#1#2{%
\setbox0=\hbox{$#1\oldsqrt{#2\,}$}\dimen0=\ht0
\advance\dimen0-0.2\ht0
\setbox2=\hbox{\vrule height\ht0 depth -\dimen0}
{\box0\lower0.4pt\box2}}

\numberwithin{equation}{section}

\title{\bf{Non-perturbative constraints on the quark and ghost propagators}} 
\author{\large{Peter Lowdon} \\
\ \\
\textit{\small{SLAC National Accelerator Laboratory, 2575 Sand Hill Rd, Menlo Park, CA 94025, USA}} \\
\textit{\small{E-mail: lowdon@slac.stanford.edu}}}
\date{}
\begin{document}
\begin{flushright} SLAC-PUB-17181 \end{flushright}
\vspace{15mm} 
{\let\newpage\relax\maketitle}
\setcounter{page}{1}
\pagestyle{plain}

\abstract 

\noindent 
In QCD both the quark and ghost propagators are important for governing the non-perturbative dynamics of the theory. It turns out that the dynamical properties of the quark and ghost fields impose non-perturbative constraints on the analytic structure of these propagators. In this work we explicitly derive these constraints. In doing so we establish that the corresponding spectral densities include components which are multiples of discrete mass terms, and that the propagators are permitted to contain singular contributions involving derivatives of $\delta(p)$, both of which are particularly relevant in the context of confinement.

\newpage

\section{Introduction}

The non-perturbative behaviour of propagators involving coloured fields enters into many important areas of quantum chromodynamics (QCD), including the dynamics of quark-gluon plasma~\cite{Cassing_Bratkovskaya09,Maas13} and the nature of confinement itself~\cite{Kugo_Ojima79,Nakanishi_Ojima90,Alkofer_vonSmekal01, Alkofer_Greensite07,Lowdon16,Lowdon17_1}. Nevertheless, the overall structure of these objects remains largely unknown. In order to gain a better understanding of the general characteristics of these objects, one requires a framework in which one can probe the non-perturbative regime. In the literature, many of the new insights into the structure of QCD propagators have come from non-perturbative numerical approaches~\cite{Alkofer_vonSmekal01,Alkofer_Detmold_Fischer_Maris04,Cucchieri_Mendes_Taurines05,Cucchieri_Mendes08,Oliveira_Silva09,Strauss_Fischer_Kellermann12, Dudal_Oliveira_Silva14}. Now whilst these approaches provide a powerful way to calculate certain aspects of propagators, they necessarily contain uncertainties due to the approximations that are required in order to carry out the calculations\footnote{In the case of the solutions of the Schwinger-Dyson equations, uncertainties arise for example due to the choice of truncation scheme employed in order to consistently solve the equations.}. In Refs.~\cite{Lowdon17_1} and~\cite{Lowdon18_1} an alternative approach was developed in order to establish the most general structural form of the gluon propagator. This approach inovlved applying a local quantum field theory (LQFT) framework, which is constructed via the assertion of a series of physically motivated axioms~\cite{Nakanishi_Ojima90,Streater_Wightman64,Haag96,Strocchi13,Bogolubov_Logunov_Oksak90}. Since these axioms are assumed to hold independently of the coupling regime, this enables genuine non-perturbative characteristics to be derived in a purely analytic manner. \\

\noindent
An important feature of gauge theories such as QCD is that the gauge symmetry provides an obstacle to the locality of the theory\footnote{By locality we mean that the fields in the theory are local fields, and therefore commute or anti-commute (depending on their spin properties) for space-like separations~\cite{Strocchi13}.}. In order to construct a consistent quantised theory one is left with two options: either one allows non-local fields, or one preserves locality. A general feature of local quantisations is that additional degrees of freedom are introduced into the theory, resulting in a space of states with an indefinite inner product. The prototypical example is the Becchi-Rouet-Stora-Tyutin (BRST) quantisation of QCD, where the space of states $\mathcal{V}_{\text{QCD}}$ contains negative-norm ghost states. In this case the physical states $\mathcal{V}_{\text{phys}} \subset \mathcal{V}_{\text{QCD}}$ correspond to those that are annihilated by the BRST charge~\cite{Nakanishi_Ojima90}. Although many of the generic features of positive-definite inner product QFTs are preserved in BRST quantised QCD, it turns out that the existence of an indefinite inner product can lead to significant changes in the structure of the propagators\footnote{LQFTs defined with an indefinite inner product space of states can be described using a modified version of the standard QFT axioms, which are often referred to as the \textit{Pseudo-Wightman axioms}. A more in-depth discussion of this framework can be found in Ref.~\cite{Bogolubov_Logunov_Oksak90}.}. In particular, in Ref.~\cite{Lowdon17_1} it was demonstrated that the BRST quantised gluon propagator can potentially contain singular terms involving derivatives of $\delta(p)$, a feature which is related to confinement~\cite{Lowdon16,Strocchi76,Strocchi78}. \\

\noindent
Since quark and ghost fields are the other degrees of freedom in QCD for which the propagators play a central role in governing the dynamics of the theory, it is also important to determine the structural properties of the propagators associated with these fields. With this motivation in mind, the aim of this paper is to continue the approach developed in Refs.~\cite{Lowdon17_1} and~\cite{Lowdon18_1} for the gluon propagator, and evaluate the constraints imposed on the quark and ghost propagators in BRST quantised QCD. The rest of the paper is organised as follows: in Sec.~\ref{np_gen_ferm} a local QFT approach is used to derive the general structural representation of the Lorentz covariant Dirac fermion correlator and propagator, and these representations are then used together with the quark Schwinger-Dyson equation to constrain the quark propagator; in Sec.~\ref{np_ghost_gen} an analogous approach is applied in order to determine the overall structural form of an anti-commuting ghost correlator and propagator, and the subsequent constraints imposed by the Schwinger-Dyson equation on the QCD ghost propagator; and finally in Sec.~\ref{concl} the main findings are summarised.

\section{Non-perturbative constraints on the quark propagator}
\label{np_gen_ferm}

In order to derive the structural form of the quark propagator in QCD one must first determine the general properties of an arbitrary Dirac fermion correlator and propagator. These properties will be discussed in the proceeding sections.

\subsection{The Dirac fermion correlator}
\label{ferm_corr_sect}

A central feature of local formulations of QFT is that correlators $\langle 0 | \phi_{1}(x_{1})\phi_{2}(x_{2}) |0\rangle$ and their Fourier transforms $\widehat{T}_{(1,2)}(p)$ are distributions\footnote{More specifically, they are assumed to belong to the class of tempered distributions $\mathcal{S}'(\mathbb{R}^{1,3})$~\cite{Streater_Wightman64}.}. Due to the Lorentz transformation properties of the fields $\phi_{1}$ and $\phi_{2}$ it follows that $\widehat{T}_{(1,2)}(p)$ can be decomposed in the following manner:
\begin{align}
\widehat{T}_{(1,2)}(p) = \sum_{\alpha=1}^{\mathscr{N}}Q_{\alpha}(p) \, \widehat{T}_{\alpha (1,2)}(p),
\label{decomp_cov}
\end{align}    
where $\widehat{T}_{\alpha (1,2)}(p)$ are Lorentz invariant distributions, and $Q_{\alpha}(p)$ are polynomial functions of $p$ carrying the same Lorentz index structure as $\phi_{1}$ and $\phi_{2}$~\cite{Bogolubov_Logunov_Oksak90}. The first case of interest in this paper is where $\phi_{1}=\psi$ and $\phi_{2}=\overbar{\psi}$ are Dirac spinor and conjugate spinor fields respectively. In this instance there are two possible Lorentz covariant polynomials: $Q_{1}(p) = \mathbb{I}$ and $Q_{2}(p) = \gamma^{\mu}p_{\mu} = \slashed{p}$, where the spinor indices have been suppressed. It follows from Eq.~(\ref{decomp_cov}) that the momentum space fermion correlator can then be written 
\begin{align}
\widehat{S}(p) = \mathcal{F}\left[\langle 0| \psi(x)\overbar{\psi}(y)|0\rangle \right] =  \mathbb{I} \, \widehat{S}_{1}(p) + \slashed{p} \, \widehat{S}_{2}(p).
\label{fermion_decomp}
\end{align}  
\ \\
\noindent
As in the case of the vector correlator~\cite{Lowdon17_1}, the Lorentz invariant distributions $\widehat{S}_{1}(p)$ and $\widehat{S}_{2}(p)$ are restricted to have support in the closed forward light cone $\overbar{V}^{+}$, and therefore have the following spectral representation~\cite{Bogolubov_Logunov_Oksak90}:   
\begin{align}
\widehat{S}_{i}(p) =  P_{i}(\partial^{2})\delta(p) + \int_{0}^{\infty} ds \, \theta(p^{0})\delta(p^{2}-s) \rho_{i}(s), 
\label{KL_gen_rep} 
\end{align}   
where $P_{i}(\partial^{2})$ is a polynomial of finite order in the d'Alembert operator $\partial^{2} = g_{\mu\nu}\frac{\partial}{\partial p_{\mu}}\frac{\partial}{\partial p_{\nu}}$, and $\rho_{i}(s)$ are the corresponding spectral densities\footnote{It turns out that the spectral densities $\rho_{i}(s)$ are tempered distributions in the class $\mathcal{S}'(\overbar{\mathbb{R}}_{+})$.}. The full fermion correlator therefore takes the form 
\begin{align}
\widehat{S}(p) =    \int_{0}^{\infty} ds \, \theta(p^{0})\delta(p^{2}-s) \left[ \rho_{1}(s) + \slashed{p} \rho_{2}(s) \right] + \left[P_{1}(\partial^{2}) + \slashed{p} P_{2}(\partial^{2}) \right]\delta(p).
\label{fermion_decomp_further}
\end{align}
Taking the inverse Fourier transform of this expression leads to the general representation of the position space correlator
\begin{align}
\langle 0| \psi(x)\overbar{\psi}(y)|0\rangle  =  & -\int_{0}^{\infty} \frac{ds}{2\pi} \left[ \rho_{1}(s) +\rho_{2}(s) i\slashed{\partial}  \right]iD^{(-)}(x-y;s)  \nonumber \\
&\hspace{5mm} +\frac{1}{(2\pi)^{4}}\Big[ P_{1}\left[-(x-y)^{2}\right]+i\slashed{\partial} \, P_{2}\left[-(x-y)^{2}\right] \Big],
\label{x_corr}
\end{align}
where $D^{(-)}(x-y;s)$ is the negative frequency Pauli-Jordan function~\cite{Bogolubov_Logunov_Oksak90}. Since $P_{1}$ and $P_{2}$ are complex polynomials of finite order, one can set: $P_{1} = \sum_{l=0}a_{l}\left[-(x-y)^{2}\right]^{l}$, and $P_{2} = \sum_{m=1}b_{m}\left(-(x-y)^{2}\right)^{m}$ where $a_{l},b_{m} \in \mathbb{C}$. The sum in $P_{2}$ does not include the $m=0$ term because this will not contribute due to the derivative in Eq.~(\ref{x_corr}).

\subsection{The Dirac fermion propagator}

The fermion propagator involves a time-ordered product of fields, and is defined by
\begin{align}
\langle 0|T\{ \psi(x)\overbar{\psi}(y)\}|0\rangle : = \theta(x^{0}-y^{0}) \langle 0| \psi(x)\overbar{\psi}(y)|0\rangle - \theta(y^{0}-x^{0})\langle 0| \overbar{\psi}(y)\psi(x)|0\rangle.
\label{t_ordered_ferm}
\end{align}
In order to determine the spectral representation of this propagator one must first establish the spectral representation for the correlator $\langle 0| \overbar{\psi}(y)\psi(x)|0\rangle$. Since the CPT operator $\Theta$ transforms Dirac spinor fields as: $\Theta \psi(x) \Theta^{-1} = i \gamma_{5}\psi^{\dagger}(-x)$, and the vacuum state is invariant under the action of $\Theta$, one has the following relation
\begin{align}
\langle 0| \overbar{\psi}(y)\psi(x)|0\rangle = - \gamma_{5} \langle 0|\psi(-x) \overbar{\psi}(-y)|0\rangle \gamma_{5}.
\label{CPT_ferm}
\end{align}   
Using the spectral representation of the fermion correlator in Eq.~(\ref{fermion_decomp_further}), the propagator can then be written
\begin{align}
\langle 0|T\{ \psi(x)\overbar{\psi}(y)\}|0\rangle &= \theta(x^{0}-y^{0}) \int_{0}^{\infty} ds \int \frac{d^{4}p}{(2\pi)^{4}} \, e^{-ip(x-y)}\theta(p^{0})\delta(p^{2}-s) \left[ \rho_{1}(s) + \slashed{p} \rho_{2}(s) \right] \nonumber \\
& \hspace{5mm} + \theta(x^{0}-y^{0})\int \frac{d^{4}p}{(2\pi)^{4}} \, e^{-ip(x-y)} \left[P_{1}(\partial^{2}) + \slashed{p} P_{2}(\partial^{2}) \right]\delta(p) \nonumber \\
&\hspace{5mm}+ \theta(y^{0}-x^{0}) \int_{0}^{\infty} ds \int \frac{d^{4}p}{(2\pi)^{4}} \, e^{ip(x-y)} \theta(p^{0})\delta(p^{2}-s) \left[ \rho_{1}(s) - \slashed{p} \rho_{2}(s) \right] \nonumber \\
& \hspace{5mm} + \theta(y^{0}-x^{0})\int \frac{d^{4}p}{(2\pi)^{4}} \, e^{ip(x-y)} \left[P_{1}(\partial^{2}) - \slashed{p} P_{2}(\partial^{2}) \right]\delta(p).
\label{t_ordered} 
\end{align}
In order to simplify this expression one can use the relation
\begin{align*}
i\slashed{\partial}\left[\theta(x^{0}-y^{0})e^{-ip(x-y)} + \theta(y^{0}-x^{0})e^{ip(x-y)}   \right] = & \slashed{p}\left[\theta(x^{0}-y^{0})e^{-ip(x-y)} - \theta(y^{0}-x^{0})e^{ip(x-y)}   \right] \\
& +i\gamma^{0} \, \delta(x^{0}-y^{0})\left[ e^{-ip(x-y)} - e^{ip(x-y)} \right], 
\end{align*} 
which upon substitution into Eq.~(\ref{t_ordered}) implies that the Dirac fermion propagator has the following general structure
\begin{align}
\langle 0|T\{ \psi(x)\overbar{\psi}(y)\}|0\rangle &=   -\int_{0}^{\infty} \frac{ds}{2\pi} \, \left[ \rho_{1}(s) + \rho_{2}(s) i\slashed{\partial} \right]i\Delta_{F}(x-y;s) \nonumber \\
&\hspace{15mm} +\frac{1}{(2\pi)^{4}}P_{1}\left[-(x-y)^{2}\right]+\frac{i}{(2\pi)^{4}}\slashed{\partial} P_{2}\left[-(x-y)^{2}\right],
\label{general_propagator_pos}
\end{align} 
where $\Delta_{F}(x-y;s)$ is the Green's function of the Klein-Gordon equation. The momentum space propagator $\widehat{S}_{F}(p)$ therefore has the form 
\begin{align}
\widehat{S}_{F}(p) &=   i\int_{0}^{\infty} \frac{ds}{2\pi} \, \frac{\left[ \rho_{1}(s) + \slashed{p}\rho_{2}(s) \right]}{p^{2}-s +i\epsilon}  + \left[ P_{1}(\partial^{2}) + \slashed{p} P_{2}(\partial^{2})\right]\delta(p).  
\label{general_ferm_propagator_mom}
\end{align}

\ \\

\noindent
The representations in Eqs.~(\ref{general_propagator_pos}) and~(\ref{general_ferm_propagator_mom}) follow only from the assumption that the momentum space correlators are Lorentz covariant distributions with support in the closed forward light cone. Since this assumption is a generic feature of any QFT, these representations are therefore model independent.

\subsection{The quark propagator}
\label{QCD_fermion}

Since the general spectral properties of a Dirac fermion propagator have been outlined in the previous section, one can now use the dynamical information in BRST quantised QCD to derive the model-dependent constraints on the structure of the quark propagator.

\subsubsection{General structure}
\label{gen_quark_struct}

In BRST quantised QCD the renormalised quark field $\psi^{i}$ satisfies the equation of motion
\begin{align}
(i\gamma^{\mu}\partial_{\mu} - m)\psi^{i} = -g\gamma^{\mu}A_{\mu}^{a}(x)(t^{a}\psi)^{i} = \mathcal{K}^{i},
\label{quark_eom}
\end{align}  
where $t^{a}$ is the colour group generator in the fundamental representation, $i$ is the colour index, and $g, m$ are the renormalised coupling and mass parameters. The quark fields also satisfy the equal-time anti-commutation relation
\begin{align}
\left\{\psi^{i}(x),\overbar{\psi}^{j}(y) \right\}_{x_{0}=y_{0}} = \delta^{ij}Z_{2}^{-1}\gamma^{0}\delta( \mathbf{x}-\mathbf{y}),
\label{quark_etar}
\end{align}
where $Z_{2}$ is the quark field renormalisation constant. Taking the vacuum expectation value of Eq.~(\ref{quark_etar}), and applying Eq.~(\ref{x_corr}) together with Eq.~(\ref{CPT_ferm}) gives
\begin{align}
\langle 0|\{\psi^{i}(x),\overbar{\psi}^{j}(y) \}|0\rangle_{x_{0}=y_{0}} &= -\left[\int_{0}^{\infty} \frac{ds}{2\pi}\left[\rho^{ij}_{1}(s) + \rho^{ij}_{2}(s)i\slashed{\partial}\right]iD^{(-)}(x-y;s) \right]_{x_{0}=y_{0}}  \nonumber \\
&+\left[\int_{0}^{\infty} \frac{ds}{2\pi}\left[\rho^{ij}_{1}(s) + \rho^{ij}_{2}(s)i\slashed{\partial}\right]iD^{(-)}(y-x;s)\right]_{x_{0}=y_{0}} \nonumber \\
&= -\left[\int_{0}^{\infty} \frac{ds}{2\pi}\left[\rho^{ij}_{1}(s) + \rho^{ij}_{2}(s)\left(i\gamma^{0}\partial_{0} + i\gamma^{j}\partial_{j}\right)\right]iD(x-y;s)\right]_{x_{0}=y_{0}} 
\label{IC_ferm}
\end{align} 
Using the initial conditions: $D(x-y;s)_{x_{0}=y_{0}}=0$ and $\dot{D}(x-y;s)_{x_{0}=y_{0}} = \delta(\mathbf{x}-\mathbf{y})$, and comparing with Eq.~(\ref{quark_etar}), it follows from Eq.~(\ref{IC_ferm}) that $\rho_{2}^{ij}(s)$ satisfies the spectral density constraint
\begin{align}
\int_{0}^{\infty} ds \, \rho_{2}^{ij}(s) = 2\pi \delta^{ij} Z_{2}^{-1}.
\label{q_int_rho2}
\end{align} 
In contrast to the gluon propagator case~\cite{Lowdon17_1}, the equal-time anti-commutation relation imposes an integral constraint on one of the spectral densities, not both. \\

\noindent
BRST quantised QCD has a space of states with an indefinite inner product. Among other things this implies that not all correlators are guaranteed to define positive-definite distributions~\cite{Bogolubov_Logunov_Oksak90}. In certain cases, such as correlators constructed from gauge-invariant fields, one can demonstrate though that correlators do indeed possess this property. However, since the interacting quark correlator $\langle 0|\psi(x)^{i}\overbar{\psi}^{j}(y)|0\rangle$ itself is not composed of gauge-invariant fields, nor is it related to a gauge-invariant correlator which consists of the quark field or its derivatives (like the photon correlator in QED~\cite{Lowdon17_1}), neither the state space structure nor the dynamical equations [Eqs.~(\ref{quark_eom}) and~(\ref{quark_etar})] are sufficient to rule out the possibility of terms involving derivatives of $\delta(p)$. In particular, this implies that the corresponding (momentum space) polynomial terms $P_{1}^{ij}(\partial^{2})=\sum_{l}a_{l}^{ij}(\partial^{2})^{l}$ and $P_{2}^{ij}(\partial^{2})= \sum_{m}b_{m}^{ij}(\partial^{2})^{m}$ for the quark correlator may be non-vanishing, and hence the quark propagator has the general form 
\begin{align}
\widehat{S}_{F}^{ij}(p) &=   i\int_{0}^{\infty} \frac{ds}{2\pi} \, \frac{\left[ \rho_{1}^{ij}(s) + \slashed{p}\rho_{2}^{ij}(s) \right]}{p^{2}-s +i\epsilon}  + \left[ P_{1}^{ij}(\partial^{2}) + \slashed{p} P_{2}^{ij}(\partial^{2})\right]\delta(p). 
\label{quark_propagator_mom}
\end{align} 
Although the overall analytic structure of the quark propagator has been discussed many times in the literature~\cite{Alkofer_vonSmekal01,Alkofer_Detmold_Fischer_Maris04,Lowdon15_1}, the possibility of singular terms in the quark propagator is a feature that has generally not been emphasised before, and yet could potentially be important in the context of QCD confinement. In Ref.~\cite{Lowdon16} it was established that the appearance of non-measure-defining terms in correlators, which includes derivatives of $\delta(p)$, can cause the correlation strength between the states created by the fields in these correlators to \textit{increase} with the separation of the states, a violation of the so-called \textit{cluster decomposition property}~\cite{Strocchi76,Strocchi78}. If one could demonstrate that this occurs for any correlator involving fields which create coloured states, this would imply that the corresponding states could not be measured independently of one another, which is a sufficient condition for confinement~\cite{Nakanishi_Ojima90,Roberts_Williams_Krein91}.

\subsubsection{Schwinger-Dyson equation constraints}
\label{QCD_quark_prop}

Now that the general structure of the quark propagator has been outlined, one can evaluate the further constraints that the equation of motion [Eq.~(\ref{quark_eom})] imposes. As demonstrated in Ref.~\cite{Lowdon18_1} for the gluon propagator, a direct way to determine these constraints is to derive the corresponding Schwinger-Dyson equation, and then use this to separately constrain the singular and non-singular terms in the propagator. Combining Eq.~(\ref{quark_eom}) and Eq.~(\ref{quark_etar}), together with the definition of the Dirac fermion propagator in Eq.~(\ref{t_ordered_ferm}), one obtains the coordinate space quark Schwinger-Dyson equation 
\begin{align}
(i\gamma^{\mu}\partial_{\mu} - m)\langle 0|T\{ \psi^{i}(x)\overbar{\psi}^{j}(y)\}|0\rangle = i\delta^{ij}Z_{2}^{-1}\delta(x-y) + \langle 0|T\{ \mathcal{K}^{i}(x)\overbar{\psi}^{j}(y)\}|0\rangle,
\label{quark_sde_x}
\end{align} 
which in momentum space has the form
\begin{align}
(\slashed{p} - m)\widehat{S}^{ij}_{F}(p) = i\delta^{ij}Z_{2}^{-1} + \widehat{K}^{ij}(p),
\label{quark_sde_p}
\end{align}
where $\widehat{K}^{ij}(p):= \mathcal{F}\left[\langle 0|T\{ \mathcal{K}^{i}(x)\overbar{\psi}^{j}(y)\}|0\rangle  \right]$. Since $\mathcal{K}^{i}(x) := -g\gamma^{\mu}A_{\mu}^{a}(x)[t^{a}\psi(x)]^{i}$ transforms as a Dirac spinor, $\widehat{K}^{ij}(p)$ has an analogous spectral representation to $\widehat{S}^{ij}_{F}(p)$ 
\begin{align}
\widehat{K}^{ij}(p) &=   i\int_{0}^{\infty} \frac{ds}{2\pi} \, \frac{\left[ \widetilde{\rho}_{1}^{ij}(s) + \slashed{p}\widetilde{\rho}_{2}^{ij}(s) \right]}{p^{2}-s +i\epsilon}  + \left[ \widetilde{P}_{1}^{ij}(\partial^{2}) + \slashed{p} \widetilde{P}_{2}^{ij}(\partial^{2})\right]\delta(p).  
\label{QCD_K_propagator_mom}
\end{align}
Inserting Eqs.~(\ref{quark_propagator_mom}) and~(\ref{QCD_K_propagator_mom}) into Eq.~(\ref{quark_sde_p}), and separately equating the terms involving derivatives of $\delta(p)$ which have support solely at $p=0$, and the terms with support outside of $p=0$, one obtains the equalities
\begin{align}
&\left(\slashed{p} - m \right)\left[ P_{1}^{ij}(\partial^{2}) + \slashed{p} P_{2}^{ij}(\partial^{2})\right]\delta(p)  = \left[ \widetilde{P}_{1}^{ij}(\partial^{2}) + \slashed{p} \widetilde{P}_{2}^{ij}(\partial^{2})\right]\delta(p),
\label{q_constr_2} \\
&\left(\slashed{p} - m \right)\left[i\int_{0}^{\infty} \frac{ds}{2\pi} \, \frac{\left[ \rho_{1}^{ij}(s) + \slashed{p}\rho_{2}^{ij}(s) \right]}{p^{2}-s +i\epsilon} \right] = i\delta^{ij}Z_{2}^{-1} + \left[i\int_{0}^{\infty} \frac{ds}{2\pi} \, \frac{\left[ \widetilde{\rho}_{1}^{ij}(s) + \slashed{p}\widetilde{\rho}_{2}^{ij}(s) \right]}{p^{2}-s +i\epsilon} \right]. \label{q_constr_1}
\end{align} 
In order to determine the relations imposed by Eq.~(\ref{q_constr_2}), let $ \widetilde{P}_{1}^{ij}(\partial^{2})=\sum_{r}\tilde{a}_{r}^{ij}(\partial^{2})^{r}$ and $ \widetilde{P}_{2}^{ij}(\partial^{2})= \sum_{s}\tilde{b}_{s}^{ij}(\partial^{2})^{s}$ be the polynomial terms of the propagator $\widehat{K}^{ij}(p)$. By equating the terms proportional to $\slashed{p}$ and the Dirac spinor identity, one obtains the following constraints on the coefficients of $P_{1}^{ij}$ and $P_{2}^{ij}$
\begin{align}
a_{n}^{ij} &=   \frac{m^{2n}}{4^{n}(n+1)!n!}\left[ a_{0}^{ij} +  \sum_{k=0}^{n-1}\frac{4^{k}(k+1)!k!\left(m\tilde{a}_{k}^{ij}+4(k+1)(k+2)\tilde{b}_{k+1}^{ij}\right)}{m^{2(k+1)}} \right], \hspace{5mm}  n \geq 1 \label{ab_rel1}\\
b_{n}^{ij} &= \frac{m^{2n-1}}{4^{n}(n+1)!n!}\left[ a_{0}^{ij} +  \sum_{k=0}^{n-1}\frac{4^{k}(k+1)!k!\left(m\tilde{a}_{k}^{ij}+4(k+1)(k+2)\tilde{b}_{k+1}^{ij}\right)}{m^{2(k+1)}} \right] - \frac{1}{m}\tilde{b}_{n}^{ij}, \hspace{5mm} n \geq 1 
\label{ab_rel2}
\end{align}  
Eqs.~(\ref{ab_rel1}) and~(\ref{ab_rel2}) demonstrate that $a_{n}^{ij}$ and $b_{n}^{ij}$ are completely determined by $a_{0}^{ij}$ and the coefficients of the singular terms in $\widehat{K}^{ij}(p)$. In particular, these relations imply that if the quark propagator contains a $\delta(p)$ term (i.e. $a_{0}^{ij} \neq 0$), or singular terms are present in the propagator $\widehat{K}^{ij}(p)$, this is sufficient to ensure that the quark propagator must contain terms involving derivatives of $\delta(p)$. In contrast, the coefficients of terms involving derivatives of $\delta(p)$ in the gluon propagator are not affected by the presence or absense of $\delta(p)$ terms~\cite{Lowdon18_1}.   \\

\noindent
As with Eq.~(\ref{q_constr_2}) one can perform the same matching procedure for Eq.~(\ref{q_constr_1}), and in doing so one obtains the following equalities
\begin{align}
&i\int_{0}^{\infty} \frac{ds}{2\pi}\rho_{2}^{ij}(s) +  i\int_{0}^{\infty} \frac{ds}{2\pi} \, \frac{\left[ s\rho_{2}^{ij}(s) -m\rho_{1}^{ij}(s) \right]}{p^{2}-s +i\epsilon} = i\delta^{ij} Z_{2}^{-1} + i\int_{0}^{\infty} \frac{ds}{2\pi} \, \frac{\widetilde{\rho}_{1}^{ij}(s)}{p^{2}-s +i\epsilon},  \label{q_constr_1_2} \\
&i\int_{0}^{\infty} \frac{ds}{2\pi} \, \frac{\left[ \rho_{1}^{ij}(s) -m\rho_{2}^{ij}(s) \right]}{p^{2}-s +i\epsilon} = i\int_{0}^{\infty} \frac{ds}{2\pi} \, \frac{\widetilde{\rho}_{2}^{ij}(s)}{p^{2}-s +i\epsilon}.
\label{q_constr_2_2}
\end{align} 
Using the fact that $\rho_{2}^{ij}(s)$ satisfies the integral condition in Eq.~(\ref{q_int_rho2}), Eqs.~(\ref{q_constr_1_2}) and~(\ref{q_constr_2_2}) imply the spectral density constraints
\begin{align}
&s\rho_{2}^{ij}(s) -m\rho_{1}^{ij}(s) = \widetilde{\rho}_{1}^{ij}(s), \\
&\rho_{1}^{ij}(s) -m\rho_{2}^{ij}(s) = \widetilde{\rho}_{2}^{ij}(s),
\label{rho1_constr_q}
\end{align}
which can be rewritten in the form
\begin{align}
&\left(s-m^{2}\right) \rho_{1}^{ij}(s) = m \widetilde{\rho}_{1}^{ij}(s) + s\widetilde{\rho}_{2}^{ij}(s), \\
&\left(s-m^{2}\right) \rho_{2}^{ij}(s) =  \widetilde{\rho}_{1}^{ij}(s) + m\widetilde{\rho}_{2}^{ij}(s).
\end{align}
As with the spectral densities of the gluon propagator, these distributional equations can be explicitly solved~\cite{Bogolubov_Logunov_Oksak90}, and have the following general solutions 
\begin{align}
\rho_{1}^{ij}(s) = A^{ij}_{1} \delta(s-m^{2}) + \kappa_{1}^{ij}(s), \\
\rho_{2}^{ij}(s) = A^{ij}_{2} \delta(s-m^{2}) + \kappa_{2}^{ij}(s),
\end{align}
where the components $\kappa_{1}^{ij}(s)$ and $\kappa_{2}^{ij}(s)$ are particular solutions which satisfy the relations $\left(s-m^{2}\right)\kappa_{1}^{ij}(s)= m \widetilde{\rho}_{1}^{ij}(s) + s\widetilde{\rho}_{2}^{ij}(s)$ and $\left(s-m^{2}\right) \kappa_{2}^{ij}(s) =  \widetilde{\rho}_{1}^{ij}(s) + m\widetilde{\rho}_{2}^{ij}(s)$ respectively. Therefore, $\kappa_{1}^{ij}(s)$ and $\kappa_{2}^{ij}(s)$ are completely determined by the spectral densities of $\widehat{K}^{ij}(p)$. \\

\noindent
In order to fix the coefficients $A^{ij}_{1}$ and $A^{ij}_{2}$, one must use the integral constraints on the various spectral densities. In addition to Eq.~(\ref{q_int_rho2}), it turns out that $\widetilde{\rho}_{2}^{ij}(s)$ satisfies the sum rule
\begin{align}
\int_{0}^{\infty} ds \, \widetilde{\rho}_{2}^{ij}(s) = 0.
\label{tilde_rho2}
\end{align}
This sum rule is derived from the equal-time restricted anti-commutator correlator relation 
\begin{align}
\langle 0|\left\{ \mathcal{K}^{i}(x),\overbar{\psi}^{j}(y) \right\} |0\rangle_{x_{0}=y_{0}}  =0,
\end{align}
which itself follows from Eq.~(\ref{quark_etar}) and the fact that the gluon field $A_{\mu}^{a}$ has a vanishing vacuum expectation value. Combining Eqs.~(\ref{q_int_rho2}) and~(\ref{tilde_rho2}) together with Eq.~(\ref{rho1_constr_q}), finally gives  
\begin{align}
&\rho_{1}^{ij}(s) = \left[2\pi m \, \delta^{ij} Z_{2}^{-1} - \int d\tilde{s} \, \kappa_{1}^{ij}(\tilde{s})    \right] \delta(s-m^{2}) + \kappa_{1}^{ij}(s), \\
&\rho_{2}^{ij}(s) =\left[2\pi \delta^{ij} Z_{2}^{-1} - \int d\tilde{s} \, \kappa_{2}^{ij}(\tilde{s})   \right]  \delta(s-m^{2}) + \kappa_{2}^{ij}(s).
\end{align}
These equalities explicitly demonstrate that the quark spectral densities both contain a discrete mass component. However, in contrast to the case of the gluon propagator~\cite{Lowdon18_1}, the coefficients in front of these components are not completely constrained, and depend explicitly on the integrals of $\kappa_{1}^{ij}(s)$ and $\kappa_{2}^{ij}(s)$. It is therefore not as clear-cut as to whether these mass components are actually present or absent in specific gauges.

\section{Non-perturbative constraints on the ghost propagator}
\label{np_ghost_gen}

As in the case of the quark propagator in Sec.~\ref{np_gen_ferm}, before deriving the general structural form of the ghost propagator in QCD, one must first determine the properties of an arbitrary ghost correlator and propagator.

\subsection{The ghost correlator and propagator}
\label{ghost_corr_sect}

Ghost $C^{a}$ and anti-ghost $\overbar{C}^{a}$ fields are anti-commuting scalar fields. From the general analysis in Sec.~\ref{ferm_corr_sect} it follows that the momentum space ghost correlator can be written 
\begin{align}
\widehat{G}^{ab}(p) = \mathcal{F}\left[\langle 0| C^{a}(x)\overbar{C}^{b}(y)|0\rangle \right] =  P^{ab}_{C}(\partial^{2})\delta(p) + \int_{0}^{\infty} ds \, \theta(p^{0})\delta(p^{2}-s) \rho_{C}^{ab}(s), 
\label{ghost_decomp}
\end{align}  
where $P^{ab}_{C}=\sum_{n}g_{n}^{ab}\left[-(x-y)^{2}\right]^{n}$ is a polynomial of finite order. Taking the inverse Fourier transform of this expression then leads to the following general representation of the position space correlator:
\begin{align}
\langle 0| C^{a}(x)\overbar{C}^{b}(y)|0\rangle  =   -\int_{0}^{\infty} \frac{ds}{2\pi}  \rho_{C}^{ab}(s)iD^{(-)}(x-y;s)   +\frac{1}{(2\pi)^{4}} P_{C}^{ab}\left[-(x-y)^{2}\right].
\label{x_corr_ghost}
\end{align}
The corresponding propagator for a general ghost field is defined by
\begin{align}
\langle 0|T\{ C^{a}(x)\overbar{C}^{b}(y)\}|0\rangle : = \theta(x^{0}-y^{0}) \langle 0| C^{a}(x)\overbar{C}^{b}(y)|0\rangle - \theta(y^{0}-x^{0})\langle 0| \overbar{C}^{b}(y)C^{a}(x)|0\rangle,
\label{t_ordered_ghost}
\end{align}
where the minus sign arises because the fields are anti-commuting. Unlike the fermion propagator, CPT symmetry cannot be used to directly relate the ghost $\langle 0| C^{a}(x)\overbar{C}^{b}(y)|0\rangle$ and anti-ghost $\langle 0| \overbar{C}^{b}(y)C^{a}(x)|0\rangle$ correlators with one another. The reason for this stems from the fact that ghost and anti-ghost fields transform as Lorentz scalars but are defined to be anti-commuting, which causes a violation of the CPT theorem~\cite{Bogolubov_Logunov_Oksak90}. The CPT operator $\Theta$ therefore does not transform the ghost and anti-ghost fields into one another, and thus the corresponding correlators must be treated independently. Nevertheless, since the anti-ghost correlator has the same distributional properties as the ghost correlator, the spectral representation has the same general structure 
\begin{align}
\mathcal{F}\left[\langle 0| \overbar{C}^{a}(y)C^{b}(x)|0\rangle \right] =  P^{ab}_{\overbar{C}}(\partial^{2})\delta(p) + \int_{0}^{\infty} ds \, \theta(p^{0})\delta(p^{2}-s) \rho_{\overbar{C}}^{ab}(s),
\label{anti_ghost_decomp}
\end{align} 
where $P^{ab}_{\overbar{C}}$ is some finite order polynomial, and $\rho_{\overbar{C}}^{ab}(s)$ is the anti-ghost spectral density. Moreover, since one defines the ghost and anti-ghost fields to be hermitian: $C^{a}(x)^{\dagger}=C^{a}(x)$, $\overbar{C}^{a}(x)^{\dagger}=\overbar{C}^{a}(x)$~\cite{Nakanishi_Ojima90}, applying the hermitian operator to Eq.~(\ref{anti_ghost_decomp}) and comparing this with Eq.~(\ref{x_corr_ghost}) implies the relations
\begin{align}
\rho_{C}^{ab}(s) = \left[\rho_{\overbar{C}}^{ba}(s)\right]^{\dagger}, \hspace{5mm} P^{ab}_{C} = \left[P^{ba}_{\overbar{C}}\right]^{\dagger}.
\label{herm_ghost}
\end{align}
Although the violation of CPT symmetry prevents the ghost and anti-ghost correlators being linearly related, the hermitian property of the fields implies that the ghost and anti-ghost spectral densities are hermitian conjugates of one another. \\

\noindent
Combining Eqs.~(\ref{x_corr_ghost}) and~(\ref{anti_ghost_decomp}) together with the definition of the propagator in Eq.~(\ref{t_ordered_ghost}), the ghost propagator takes the following form
\begin{align}
\langle 0|T\{ C^{a}(x)\overbar{C}^{b}(y)\}|0\rangle  = &- \int_{0}^{\infty} \frac{ds}{2\pi}  \rho_{C}^{ab}(s) \, i\Delta_{F}(x-y;s) +  \int \frac{d^{4}p}{(2\pi)^{4}}e^{-ip(x-y)}P^{ab}_{C}(\partial^{2})\delta(p) \nonumber \\
& - \theta(y^{0}-x^{0})\int_{0}^{\infty} \frac{ds}{2\pi}  \left[\rho_{C}^{ab}(s) + \rho_{\overbar{C}}^{ba}(s)      \right]iD^{(+)}(x-y;s)  \nonumber \\
& - \theta(y^{0}-x^{0})\int \frac{d^{4}p}{(2\pi)^{4}}e^{-ip(x-y)}\left[ P^{ab}_{C}(\partial^{2}) + P^{ba}_{\overbar{C}}(\partial^{2}) \right]\delta(p).
\label{ghost_prop}
\end{align}
Since the spectral densities $\rho_{C}^{ab}(s)$ and $\rho_{\overbar{C}}^{ab}(s)$ are only related via hermitian conjugation, one cannot simplify this expression further without additional constraints.

\subsection{The ghost propagator in QCD}

Using the general spectral properties outlined in the previous section, one can now use the dynamical characteristics of BRST quantised QCD to derive explicit constraints on the structure of the QCD ghost propagator.

\subsubsection{General structure}

In BRST quantised QCD the renormalised ghost field $C^{a}$ satisfies the equation of motion 
\begin{align}
&\partial^{2}C^{a}= -igf^{abc}\partial^{\nu}(A_{\nu}^{b}C^{c}) = \mathcal{L}^{a}, \label{gh_eom}
\end{align}
together with the equal-time anti-commutation relations
\begin{align}
&\{ C^{a}(x),\overbar{C}^{b}(y) \}_{x_{0}=y_{0}}=0, \label{gh_etar1} \\
&\{ \dot{C}^{a}(x), \overbar{C}^{b}(y)\}_{x_{0}=y_{0}} = \delta^{ab}\widetilde{Z}_{3}^{-1}\delta(\mathbf{x}-\mathbf{y}), \label{gh_etar2}
\end{align}
where $\widetilde{Z}_{3}$ is the ghost renormalisation constant. Taking the vacuum expectation values of Eqs.~(\ref{gh_etar1}) and~(\ref{gh_etar2}), and applying Eq.~(\ref{x_corr_ghost}), one obtains the conditions
\begin{align}
&P_{C}^{ab} = - P_{\overbar{C}}^{ba}, \\
&\int_{0}^{\infty} ds \, \rho_{C}^{ab}(s) = 2\pi i \delta^{ab}\widetilde{Z}_{3}^{-1},  \label{rho_c_sum}\\
&\left[\int_{0}^{\infty}ds \left[ \rho_{C}^{ab}(s) + \rho_{\overbar{C}}^{ba}(s)\right] D^{(+)}(x-y;s)\right]_{x_{0}=y_{0}} = 0, \label{IC1}\\
&\left[\int_{0}^{\infty}ds \left[ \rho_{C}^{ab}(s) + \rho_{\overbar{C}}^{ba}(s)\right] \dot{D}^{(+)}(x-y;s)\right]_{x_{0}=y_{0}} = 0. \label{IC2}
\end{align}
Since $\int_{0}^{\infty}ds \left[ \rho_{C}^{ab}(s) + \rho_{\overbar{C}}^{ba}(s)\right] D^{(+)}(x-y;s)$ satisfies the Klein-Gordon equation, the solution of this distribution for unequal times is uniquely determined by the initial conditions in Eqs.~(\ref{IC1}) and~(\ref{IC2})~\cite{Bogolubov_Logunov_Oksak90}. Furthermore, since this solution depends linearly on the initial conditions, both of which are vanishing, this implies  
\begin{align}
\int_{0}^{\infty}ds \left[ \rho_{C}^{ab}(s) + \rho_{\overbar{C}}^{ba}(s)\right] D^{(+)}(x-y;s) =0.
\end{align}
Combining all of these constraints together with the representation in Eq.~(\ref{ghost_prop}), the non-perturbative ghost propagator can then be written
\begin{align}
\langle 0|T\{ C^{a}(x)\overbar{C}^{b}(y)\}|0\rangle  = &- \int_{0}^{\infty} \frac{ds}{2\pi}  \rho_{C}^{ab}(s) \, i\Delta_{F}(x-y;s) +  \int \frac{d^{4}p}{(2\pi)^{4}}e^{-ip(x-y)}P^{ab}_{C}(\partial^{2})\delta(p),
\label{ghost_prop_QCD}
\end{align}   
which in momentum space is given by
\begin{align}
\widehat{G}_{F}^{ab}(p) &=   i\int_{0}^{\infty} \frac{ds}{2\pi} \, \frac{ \rho_{C}^{ab}(s) }{p^{2}-s +i\epsilon}  + P_{C}^{ab}(\partial^{2})\delta(p).
\label{ghost_prop_mom_QCD}
\end{align}
Since the ghost field transforms as a Lorenz scalar it is not surprising that the propagator has the same overall structure as a scalar propagator. However, unlike with standard commuting scalar fields, the structure in Eq.~(\ref{ghost_prop_mom_QCD}) depends crucially on the equal-time anti-commutation relations in Eqs.~(\ref{gh_etar1}) and~(\ref{gh_etar2}). Eq.~(\ref{rho_c_sum}) is equivalent to the sum rule satisfied by the gluon spectral density, which is proportional to the inverse of the gluon field renormalisation constant~\cite{Lowdon18_1}. Since $\widetilde{Z}_{3}^{-1}$ similarly vanishes in Landau gauge, the ghost spectral density therefore also obeys the Oehme-Zimmermann superconvergence relation~\cite{Oehme_Zimmermann80_1,Oehme_Zimmermann80_2}. As in the case of the interacting quark propagator, the potential appearance of singular terms in the ghost propagator is relevant for understanding confinement. In fact, this is particularly true for the ghost propagator, since the infrared behaviour of this object plays a central role in the Kugo-Ojima confinement criterion~\cite{Kugo_Ojima79,Nakanishi_Ojima90,Alkofer_vonSmekal01}.

\subsubsection{Schwinger-Dyson equation constraints}

In an analogous manner to Sec.~\ref{QCD_quark_prop}, one can determine the further conditions that the equation of motion [Eq.~(\ref{gh_eom})] imposes on the structure of the ghost propagator by deriving the form of the Schwinger-Dyson equation, and then using this to separately constrain the singular and non-singular terms in the propagator. Combining Eqs.~(\ref{gh_eom}),~(\ref{gh_etar1}) and~(\ref{gh_etar2}) together with the general definition of a ghost propagator in Eq.~(\ref{t_ordered_ghost}), one obtains the coordinate space Schwinger-Dyson equation
\begin{align}
\partial^{2}\langle 0|T\{ C^{a}(x)\overbar{C}^{b}(y)\}|0\rangle = \delta^{ab}\widetilde{Z}_{3}^{-1}\delta(x-y) + \langle 0|T\{ \mathcal{L}^{a}(x)\overbar{C}^{b}(y)\}|0\rangle,
\label{gh_sde_x}
\end{align} 
which in momentum space is given by
\begin{align}
-p^{2}  G^{ab}_{F}(p) = \delta^{ab}\widetilde{Z}_{3}^{-1} + L^{ab}(p),
\label{gh_sde_p}
\end{align} 
where $L^{ab}(p) = \mathcal{F}\left[\langle 0|T\{ \mathcal{L}^{a}(x)\overbar{C}^{b}(y)\}|0\rangle\right]$. Since $\mathcal{L}^{a}$ has the same Lorentz transformation properties as $C^{a}$, it follows that $L^{ab}(p)$ has an analogous spectral representation to Eq.~(\ref{ghost_prop}). Moreover, because one has the following equal-time restricted anti-commutator correlator relations\footnote{These relations follow from Eqs.~(\ref{gh_etar1}) and~(\ref{gh_etar2}), together with the fact that QCD fields have vanishing vacuum expectation values.} 
\begin{align}
\langle 0|\{ \mathcal{L}^{a}(x),\overbar{C}^{b}(y) \}|0\rangle_{x_{0}=y_{0}}=0, \hspace{5mm} \langle 0|\{ \dot{\mathcal{L}}^{a}(x), \overbar{C}^{b}(y)\}|0\rangle_{x_{0}=y_{0}} = 0, \label{L_etar}
\end{align}
the spectral representation of $L^{ab}(p)$ can be written in the same manner as for the QCD ghost propagator
\begin{align}
L^{ab}(p)  &=   i\int_{0}^{\infty} \frac{ds}{2\pi} \, \frac{\widetilde{\rho}_{C}^{ab}(s)}{p^{2}-s +i\epsilon}  + \widetilde{P}_{C}^{ab}(\partial^{2})\delta(p),  
\label{QCD_L1_propagator_mom}
\end{align} 
where now the corresponding spectral density $\widetilde{\rho}_{C}^{ab}(s)$ instead satisfies the constraint
\begin{align}
\int_{0}^{\infty} ds \, \widetilde{\rho}_{C}^{ab}(s) = 0.
\label{tilde_rho_c}
\end{align}
Inserting Eqs.~(\ref{ghost_prop_mom_QCD}) and~(\ref{QCD_L1_propagator_mom}) into Eq.~(\ref{gh_sde_p}), and separately equating the terms involving derivatives of $\delta(p)$ and those with support outside of $p=0$, one obtains
\begin{align}
&-p^{2}  \left[P_{C}^{ab}(\partial^{2})\delta(p) \right] = \widetilde{P}_{C}^{ab}(\partial^{2})\delta(p), \label{constr_gh_2} \\
&-p^{2}  \left[ i\int_{0}^{\infty} \frac{ds}{2\pi} \, \frac{\rho_{C}^{ab}(s)}{p^{2}-s +i\epsilon}\right] = \delta^{ab}\widetilde{Z}_{3}^{-1} + i\int_{0}^{\infty} \frac{ds}{2\pi} \, \frac{\widetilde{\rho}_{C}^{ab}(s)}{p^{2}-s +i\epsilon}. \label{constr_gh_1}
\end{align}
It follows from Eq.~(\ref{constr_gh_2}) that the coefficients $g_{n}^{ab}$ and $\tilde{g}_{n}^{ab}$ of the polynomials $P_{C}^{ab}$ and $\widetilde{P}_{C}^{ab}$ respectively, satisfy the following constraint
\begin{align}
g_{n+1}^{ab} = -\frac{\tilde{g}_{n}^{ab}}{4(n+1)(n+2)}, \hspace{5mm} n \geq 0.
\label{sing_gh}
\end{align} 
Eq.~(\ref{sing_gh}) implies that the coefficients of the singular terms in the ghost propagator are completely fixed by the coefficients of the singular terms in $L^{ab}(p)$. Therefore, if $L^{ab}(p)$ contains either $\delta(p)$ or non-measure defining terms involving derivatives of $\delta(p)$, then this is sufficient to guarantee that the ghost propagator must contain non-measure defining terms. \\

\noindent
In order to determine the constraints imposed by Eq.~(\ref{constr_gh_1}) one can make use of the fact that this expression can be written in the form
\begin{align}
- i \int_{0}^{\infty} \frac{ds}{2\pi} \, \rho_{C}^{ab}(s) - i\int_{0}^{\infty} \frac{ds}{2\pi} \, \frac{s\rho_{C}^{ab}(s)}{p^{2}-s +i\epsilon} = \delta^{ab}\widetilde{Z}_{3}^{-1} + i\int_{0}^{\infty} \frac{ds}{2\pi} \, \frac{\widetilde{\rho}_{C}^{ab}(s)}{p^{2}-s +i\epsilon}.
\label{gh_eq} 
\end{align}
Since the ghost spectral density satisfies the sum rule in Eq.~(\ref{rho_c_sum}), the above equality therefore implies the following constraint  
\begin{align}
s\rho_{C}^{ab}(s) = - \widetilde{\rho}_{C}^{ab}(s).
\end{align}
Similarly to the quark spectral densities, one can solve this distributional equation in terms of $\rho_{C}^{ab}(s)$, and one obtains the solution
\begin{align}
\rho_{C}^{ab}(s) = A^{ab}\delta(s) + \kappa_{C}^{ab}(s),
\end{align} 
where the particular solution $\kappa_{C}^{ab}(s)$ satisfies the relation $s\kappa_{C}^{ab}(s) = - \widetilde{\rho}_{C}^{ab}(s)$. By applying the sum rule in Eq.~(\ref{rho_c_sum}), the ghost spectral density can then finally be written
\begin{align}
\rho_{C}^{ab}(s) = \left[2\pi i\delta^{ab}\widetilde{Z}_{3}^{-1} - \int_{0}^{\infty} d\tilde{s} \, \kappa_{C}^{ab}(\tilde{s})   \right] \delta(s) + \kappa_{C}^{ab}(s).
\label{ghost_rho}
\end{align}
Eq.~(\ref{ghost_rho}) demonstrates that the ghost spectral density contains a discrete massless component. Similarly to the quark spectral densities, the coefficient in front of this discrete component is not completely constrained since it depends on the integral of $\kappa_{C}^{ab}(\tilde{s})$, which itself is determined by $\widetilde{\rho}_{C}^{ab}(s)$. This feature is particularly for understanding confinement because it turns out that in order to violate the cluster decomposition property in QCD, this requires both the appearance of non-measure-defining terms in the correlators of coloured fields, such as derivatives of $\delta(p)$, and \textit{also} that the full space of states $\mathcal{V}_{\text{QCD}}$ has no mass gap~\cite{Strocchi76,Strocchi78}. This second requirement is still consistent with the possibility that the physical subspace $\mathcal{V}_{\text{phys}} \subset \mathcal{V}_{\text{QCD}}$ has a mass gap, as one would expect in QCD~\cite{Nakanishi_Ojima90}. In Landau gauge $\widetilde{Z}_{3}^{-1}$ vanishes, and therefore the only thing preventing the absence of a massless ghost pole is the non-vanishing of $\int_{0}^{\infty} d\tilde{s} \, \kappa_{C}^{ab}(\tilde{s})$. This feature is in contrast to the case of the gluon spectral density, where the coefficient of the massless component is entirely propotional to $Z_{3}^{-1}$, which vanishes in Landau gauge, and therefore prevents the appearance of a massless gluon state~\cite{Lowdon18_1}. Since $\int_{0}^{\infty} d\tilde{s} \, \kappa_{C}^{ab}(\tilde{s})$ can in principle be non-vanishing, this preserves the possibility that $\mathcal{V}_{\text{QCD}}$ has no mass gap, and that the cluster decomposition property can be violated for coloured states, which is a sufficient condition for confinement~\cite{Roberts_Williams_Krein91}.

\section{Conclusions}
\label{concl}

Although the quark and ghost propagators play an important role in QCD, the general analytic structure of these objects remains largely unknown. In this work we demonstrate that the  dynamical properties of the quark and ghost fields, and in particular their corresponding Schwinger-Dyson equations, impose non-perturbative constraints on these propagators. For the quark propagator it turns out that these constraints imply that both spectral densities necessarily contain massive components proportional to $\delta(s-m^{2})$, and that the presence of singular terms in the propagator involving derivatives of $\delta(p)$ are permitted. In the case of the ghost propagator the corresponding spectral density is constrained to contain a massless component proportional to $\delta(s)$, and the appearance of singular terms is also similarly permitted. The potential presence of a non-vanishing massless component in the ghost spectral density, and singular terms in the quark and ghost propagators, are of particular importance in the context of confinement. Besides the purely theoretical relevance of these results, these constraints could also provide important input for improving existing parametrisations of the QCD propagators.

\section*{Acknowledgements}
This work was supported by the Swiss National Science Foundation under contract P2ZHP2\_168622, and by the DOE under contract DE-AC02-76SF00515.

\renewcommand*{\cite}{\vspace*{-12mm}}

\end{document}